\documentclass[12pt]{article}
\usepackage{amssymb,graphicx}

\begin{document}

\newcommand{\sect}[1]{\setcounter{equation}{0}\section{#1}}
\renewcommand{\theequation}{\thesection.\arabic{equation}}
\newcommand{\be}{\begin{equation}}
\newcommand{\ee}{\end{equation}}
\newcommand{\bea}{\begin{eqnarray}}
\newcommand{\eea}{\end{eqnarray}}
\newcommand{\nonu}{\nonumber\\}
\newcommand{\beano}{\begin{eqnarray*}}
\newcommand{\eeano}{\end{eqnarray*}}
%%%%%%%%%%%%%%%%%%%%%%%%%
%%%%%%%%     GREQUES     %%%%%%%
%%%%%%%%%%%%%%%%%%%%%%%%%
\newcommand{\eps}{\epsilon}
\newcommand{\om}{\omega}
\newcommand{\vph}{\varphi}
\newcommand{\sig}{\sigma}
%%%%%%%%%%%%%%%%%%%%%%%%
%%%%%%%  C, R, Q, Z, N, Id  %%%%%
%%%%%%%%%%%%%%%%%%%%%%%%
%---- CORPS DES COMPLEXES
\newcommand{\CC}{\mbox{${\mathbb C}$}}
%---- CORPS DES REELS
\newcommand{\RR}{\mbox{${\mathbb R}$}}
%---- CORPS DES RATIONNELS
\newcommand{\QQ}{\mbox{${\mathbb Q}$}}
%---- GROUPE DES ENTIERS
\newcommand{\ZZ}{\mbox{${\mathbb Z}$}}
%---- NATURELS
\newcommand{\NN}{\mbox{${\mathbb N}$}}
%---- IDENTITE EN 12 PT
\newcommand{\1}{\mbox{\hspace{.0em}1\hspace{-.24em}I}}
\newcommand{\II}{\mbox{${\mathbb I}$}}
%%%%%%%%%%%%%%%%%%%%%%%%%%%%%%%%%
%%%%%%%%    DIVERS   %%%%%%%%%%%%
%%%%%%%%%%%%%%%%%%%%%%%%%%%%%%%%%
\newcommand{\prt}{\partial}
\newcommand{\und}[1]{\underline{#1}}
\newcommand{\wh}[1]{\widehat{#1}}
\newcommand{\wt}[1]{\widetilde{#1}}
\newcommand{\mb}[1]{\ \mbox{\ #1\ }\ }
\newcommand{\half}{\frac{1}{2}}
\newcommand{\noin}{\not\!\in}
\newcommand{\rhotimes}{\mbox{\raisebox{-1.2ex}{$\stackrel{\displaystyle\otimes}
{\mbox{\scriptsize{$\rho$}}}$}}}
\newcommand{\bin}[2]{{\left( {#1 \atop #2} \right)}}
\newcommand{\A}{{\cal A}}
\newcommand{\B}{{\cal B}}
\newcommand{\C}{{\cal C}}
\newcommand{\F}{{\cal F}}
\newcommand{\E}{{\cal E}}
\newcommand{\cP}{{\cal P}}
\newcommand{\R}{{\cal R}}
\newcommand{\T}{{\cal T}}
\newcommand{\W}{{\cal W}}
\newcommand{\cS}{{\cal S}}
\newcommand{\bS}{{\bf S}}
\newcommand{\cL}{{\cal L}}
\newcommand{\hlp}{{\RR}_+}
\newcommand{\hlm}{{\RR}_-}
\newcommand{\Hil}{{\cal H}}
\newcommand{\D}{{\cal D}}
\newcommand{\G}{{\cal G}}
\newcommand{\alg}{\C} 
\newcommand{\rep}{\F(\C)}
\newcommand{\trep}{\G_\beta(\C)}
\newcommand{\form}{\langle \, \cdot \, , \, \cdot \, \rangle }
\newcommand{\e}{{\rm e}}
\newcommand{\by}{{\bf y}}
\newcommand{\bp}{{\bf p}}
\newcommand{\LL}{\mbox{${\mathbb L}$}}
\newcommand{\Rp}{{R^+_{\, \, \, \, }}}
\newcommand{\Rm}{{R^-_{\, \, \, \, }}}
\newcommand{\Rpm}{{R^\pm_{\, \, \, \, }}}
\newcommand{\Tp}{{T^+_{\, \, \, \, }}}
\newcommand{\Tm}{{T^-_{\, \, \, \, }}}
\newcommand{\Tpm}{{T^\pm_{\, \, \, \, }}}
\newcommand{\baral}{\bar{\alpha}}
\newcommand{\barbt}{\bar{\beta}}
\newcommand{\supp}{{\rm supp}\, }
\newcommand{\EE}{\mbox{${\mathbb E}$}}
\newcommand{\JJ}{\mbox{${\mathbb J}$}}
\newcommand{\MM}{\mbox{${\mathbb M}$}}
\newcommand{\ct}{{\cal T}}
\newcommand{\ph}{\varphi}
\newcommand{\phd}{\widetilde{\varphi}}
\newcommand{\phl}{\varphi_{{}_L}}
\newcommand{\phr}{\varphi_{{}_R}}
\newcommand{\phpl}{\varphi_{{}_{+L}}}
\newcommand{\phpr}{\varphi_{{}_{+R}}}
\newcommand{\phml}{\varphi_{{}_{-L}}}
\newcommand{\phmr}{\varphi_{{}_{-R}}}
\newcommand{\phpml}{\varphi_{{}_{\pm L}}}
\newcommand{\phpmr}{\varphi_{{}_{\pm R}}}
\newcommand{\Ei}{\rm Ei}

%%%%%%%% THEOREMES
\newtheorem{theo}{Theorem}[section]
\newtheorem{coro}[theo]{Corollary}
\newtheorem{prop}[theo]{Proposition}
\newtheorem{defi}[theo]{Definition}
\newtheorem{conj}[theo]{Conjecture}
\newtheorem{lem}[theo]{Lemma}
\newcommand{\prf}{\underline{\it Proof.}\ }
\newcommand{\finprf}{\null \hfill {\rule{5pt}{5pt}}\\[2.1ex]\indent}

%%%%%%%%%%%%%%%%%%%%%%%
\pagestyle{empty}
\rightline{November 2005}

\vfill

\begin{center}
{\Large\bf Bosonization and Vertex Algebras with Defects}
\\[2.1em]

\bigskip

{\large
M. Mintchev$^{a}$\footnote{mintchev@df.unipi.it} 
and P. Sorba$^{b}$\footnote{sorba@lapp.in2p3.fr}}\\

\null

\noindent 

{\it $^a$ INFN and Dipartimento di Fisica, Universit\'a di
      Pisa, Largo Pontecorvo 3, 56127 Pisa, Italy\\[2.1ex]
$^b$ LAPTH, 9, Chemin de Bellevue, BP 110, F-74941 Annecy-le-Vieux
      cedex, France}
\vfill

\end{center}

\begin{abstract} 
The method of bosonization is extended to the case when a dissipationless 
point-like defect is present in space-time. Introducing the chiral components 
of a massless scalar field, interacting with the defect in two dimensions, we construct 
the associated vertex operators. The main features of the corresponding vertex 
algebra are established. As an application of this framework we solve the massless 
Thirring model with defect. We also construct the vertex representation 
of the $\widehat {sl}(2)$ Kac-Moody algebra, describing the complex  
interplay between the left and right sectors due to  
the interaction with the defect. The Sugawara form of the energy-momentum 
tensor is also explored. 

\end{abstract}
\bigskip 
\medskip 
\centerline{\it In memory of Daniel Arnaudon} 
\bigskip

\vfill
\rightline{IFUP-TH 20/2005}
\rightline{LAPTH-1120/05}
\rightline{\tt hep-th/0511162}
\newpage
\pagestyle{plain}
\setcounter{page}{1}

%%%%%%%%%%%%%%%%%%%%%%%%%%%%%%%%
\sect{Introduction}

Bosonization represents a powerful method for solving a variety of models 
\cite{Tom}-\cite{Lutt} in two space-time 
dimensions with relevant applications in both condensed 
matter physics and string theory. On the mathematical side bosonization 
is among the fundamental tools for constructing vertex algebras \cite{Kac} and in
particular,  vertex representations of affine Kac-Moody algebras \cite{Goddard:1986bp}. 
The main goal of the present paper is to  extend the framework 
of bosonization to the case when defects
(impurities) are  present in space. The subject of defects attracts 
recently much attention in different areas 
of quantum physics, including quantum mechanics \cite{A1}-\cite{Sch}, 
integrable systems \cite{Delfino:1994nr}-\cite{Bowcock:2005vs}, 
conformal and finite temperature quantum field theory 
\cite{Saleur:1998hq}-\cite{Mintchev:2004jy} 
and string theory, where branes can be considered as purely 
reflecting defects. Some interesting studies 
\cite{Bachas:2001vj, DeWolfe:2001pq} of 
both reflecting and transmitting branes (permeable conformal walls) 
are also worth mentioning. The new configurations we introduce and 
develop here concern chiral fields and vertex operators with defects. 
The study of these structures is motivated by potential physical applications to 
condensed matter physics and string theory. On the mathematical side 
our work suggests some interesting generalizations in the context 
of vertex algebras. 

We focus below on dissipationless point-like defects, showing that they preserve 
the basic ingredients of bosonization - quantum field locality and unitarity. In 
section 2 we summarize some results about the massless scalar field 
and its dual interacting with a generic defect of the above type. 
The relative vertex operators are constructed and investigated in section 3. 
We show that they generally obey anyon statistics, thus describing 
anyon fields in the presence of defects. 
Sections 4 and 5 are devoted to some applications. In section 4 
we solve the massless Thirring model with a point-like defect and 
discuss the solution. In section 5 we study some aspects of 
non-abelian bosonization with impurities. We describe there 
the vertex operator construction of the $\widehat {sl}(2)$ Kac-Moody 
algebra, focusing on the new features stemming from the presence of a 
defect. Adopting the Sugawara representation, we investigate 
also the impact of the defect on the energy-momentum tensor.  
Finally, section 6 contains our conclusions and some indications for further
developments.

\sect{General setting} 

Bosonization (see e.~g. \cite{bos}) has a long history 
dating back \cite{JW} to the earliest years 
of quantum field theory. The main building blocks are the massless 
scalar field $\ph(t,x)$ and its dual $\phd(t,x)$. Therefore, our first step 
will be to establish the basic properties of $\{\ph,\, \phd\}$ when a 
point-like dissipationless defect is present in space. Without loss of generality 
one can localize the defect at $x=0$ and consider thus the following 
equation of motion 
\be
\left (\prt_t^2 - \prt_x^2 \right )\varphi (t,x) = 0\, , \qquad x\not= 0 \, ,  
\label{eqm}
\ee
with standard initial conditions fixed by the equal-time canonical commutation 
relations 
\be
[\varphi (0,x_1)\, ,\, \varphi (0,x_2)] = 0\, , \qquad
[(\prt_t\varphi )(0,x_1)\, ,\, \varphi (0,x_2)] = -i\delta (x_1-x_2) \, .
\label{initial}
\ee 
The most general dissipationless interaction of $\ph(t,x)$ with the defect  
at $x=0$ is described \cite{A2} by the boundary condition 
\be
\left(\begin{array}{cc} \varphi (t,+0) \\ \prt_x \varphi (t,+0)\end{array}\right) = 
\left(\begin{array}{cc} a & b\\ c&d\end{array}\right)
\left(\begin{array}{cc} \varphi (t,-0) \\ \prt_x \varphi (t,-0)\end{array}\right)\, , 
\quad \forall  t\in \RR \, ,   
\label{bc}
\ee 
where 
\be 
ad -bc = 1\, , \qquad  a,...,d \in \RR \, .  
\label{parameters}
\ee 
We observe that $a$ and $d$ are dimensionless, whereas 
$b$ and $c$ have a non-trivial and opposite dimension. 

The dual field $\phd(t,x)$ also satisfies 
\be
\left (\prt_t^2 - \prt_x^2 \right )\phd (t,x) = 0\, , \qquad x\not= 0 \, ,  
\label{eqmd}
\ee
and as usual is related to $\ph(t,x)$ by 
\be 
\prt_t \phd (t,x) = - \prt_x \ph (t,x)\, , \quad  
\prt_x \phd (t,x) = - \prt_t \ph (t,x)\, , \qquad  x\not= 0 \, .
\label{dual}
\ee 

Eqs. (\ref{eqm}-\ref{dual}) have a unique solution $\{\ph\, , \phd\}$, which 
represents the basis for bosonization with a point-like defect. In this paper 
we mostly concentrate on the case when impurity bound states are absent. 
This case is characterized \cite{Mintchev:2004jy} by the following additional 
constraints on the parameters: 
\be
 \left\{\begin{array}{cc}
 \frac{a+d + \sqrt {(a-d)^2 + 4}}{2b} \geq 0 \, ,
& \quad \mbox{for $b<0$}\, ,\\[1ex]
\frac{c}{a+d} \geq 0\, ,
& \quad \mbox{for $b=0$}\, ,\\[1ex]
 \frac{a+d - \sqrt {(a-d)^2 + 4}}{2b} \geq 0\, ,
& \quad \mbox{for $b>0$}\, . \\[1ex]
\end{array} \right.
\label{nobs}
\end{equation} 
In this domain the solution $\{\ph\, , \phd\}$ can be written in the form  
\be
\ph (t,x) = \ph_+ (t,x) + \ph_- (t,x)  \,  , \qquad 
\phd (t,x) = \phd_+ (t,x) + \phd_- (t,x)  \,  , 
\label{sol1}
\ee
where 
\be
\ph_\pm (t,x) = \theta(\pm x) \int_{-\infty}^{+\infty} \frac{dk}{2\pi \sqrt
{2|k|}}
\left[a^{\ast \pm}(k) \e^{i|k|t-ikx} +
a_\pm (k) \e^{-i|k|t+ikx}\right ] \,  , 
\label{sol2}
\ee
\be
\phd_\pm (t,x) = \theta(\pm x) \int_{-\infty}^{+\infty} 
\frac{dk\, \varepsilon (k)}{2\pi \sqrt {2|k|}} 
\left[a^{\ast \pm}(k) \e^{i|k|t-ikx} +
a_\pm (k) \e^{-i|k|t+ikx}\right ] \,  . 
\label{sol3}
\ee 
These expressions have the familiar form of superpositions of creation 
$a^{\ast \pm}(k)$ and annihilation $a_\pm (k)$ operators. The interaction 
with the impurity deforms \cite{Mintchev:2002zd} 
only their commutation relations, which read now 
\bea
&a_{\xi_1}(k_1)\, a_{\xi_2}(k_2) -  a_{\xi_2}(k_2)\, a_{\xi_1}(k_1) = 0\,  ,
\label{ccr1} \\
&a^{\ast \xi_1}(k_1)\, a^{\ast \xi_2}(k_2) - a^{\ast \xi_2}(k_2)\,
a^{\ast \xi_1}(k_1) = 0\,  ,
\label{ccr2} \\
&a_{\xi_1}(k_1)\, a^{\ast \xi_2}(k_2) - a^{\ast \xi_2}(k_2)\,
a_{\xi_1}(k_1) = \nonumber \\
&\left [\delta_{\xi_1}^{\xi_2} + T_{\xi_1}^{\xi_2}(k_1)\right ] 2\pi
\delta(k_1-k_2)\, {\bf 1} +
R_{\xi_1}^{\xi_2}(k_1) 2\pi \delta(k_1+k_2)\, {\bf 1}\,  , 
\label{ccr3}
\eea 
where 
\bea
R_+^+(k) = \frac{bk^2 + i(a-d)k + c}{bk^2 + i(a+d)k - c} \, , \qquad 
R_-^-(k) = \frac{bk^2 + i(a-d)k + c}{bk^2 - i(a+d)k - c} \, ,
\label{rcoef} \\
T_+^-(k) = \frac{2ik}{bk^2 + i(a+d)k - c}\, , \qquad 
T_-^+(k) = \frac{-2ik}{bk^2 - i(a+d)k - c}\, , 
\label{tcoef}
\eea 
are the {\it reflection} and {\it transmission coefficients} from the impurity.  
The associated {\it reflection} and {\it transmission matrices} 
\be 
R(k) = \left(\begin{array}{cc} R_+^+(k) & 0\\0 & R_-^-(k)\end{array}\right)\, , 
\qquad 
T(k) =\left(\begin{array}{cc} 0 & T_+^-(k)\\T_-^+(k) & 0\end{array}\right)\, , 
\label{rtmat}
\ee
satisfy hermitian analyticity 
\be 
R(k)^\dagger = R(-k)\,  , \qquad  T(k)^\dagger = T(k)\,  , 
\label{hanal}
\ee 
and unitarity 
\bea
T(k) T(k) + R(k) R(-k)  = \II \, , 
\label{unit1} \\
T(k) R(k) +  R(k) T(-k) = 0 \,  . 
\label{unit2}
\eea 

The exchange relations (\ref{ccr1}-\ref{ccr3}) deserve some comments. 
We observe first of all that (\ref{ccr1}-\ref{ccr3}) preserve the conventional 
initial conditions (\ref{initial}). In a slightly more general form 
the relations (\ref{ccr1}-\ref{ccr3}) appeared for the first time
\cite{Mintchev:2002zd, Mintchev:2003ue} in the context of  integrable 
models with impurities. The associative algebra generated by 
$\{a^{\ast \pm}(k),\,  a_\pm (k),\, {\bf 1}\}$, satisfying (\ref{ccr1}-\ref{ccr3}) 
and the constraints 
\bea
a_\xi(k) &=& T_\xi^\eta (k) a_\eta (k) + R_\xi^\eta (k) a_\eta (-k) \, ,
\label{c1} \\
a^{\ast \xi}(k) &=& a^{\ast \eta}(k) T_\eta^\xi (k) +
a^{\ast \eta}(-k) R_\eta^\xi (-k)\, ,  
\label{constraints}
\eea 
has been called reflection-transmission (RT) algebra because 
it translates the analytic boundary conditions (\ref{bc}) in algebraic terms, 
directly related to the physical reflection and transmission amplitudes 
(\ref{rcoef}, \ref{tcoef}). For this reason RT algebras represent a 
natural and universal tool for studying QFT with defects 
\cite{Caudrelier:2004gd,Caudrelier:2004xy,Mintchev:2004jy,Fichera:2005fp} 
and it is not at all surprising  that they appear also in the 
process of bosonization with impurities. 

The derivation of the correlation functions of $\{\ph\, , \phd\}$ in the Fock 
representation \cite{Mintchev:2003ue} 
of the RT algebra (\ref{ccr1}-\ref{ccr3}) is straightforward. 
It is convenient to change basis introducing the right and left chiral fields 
\be 
\phr(t,x)=\ph(t,x)+\phd(t,x)\, , \qquad 
\phl(t,x)=\ph(t,x)-\phd(t,x)\, . 
\label{rlbasis}
\ee 
Inserting (\ref{sol1}-\ref{sol3}) in (\ref{rlbasis}) one gets 
\bea
\phr(t,x) = \theta(x) \phpr (t-x) + \theta(-x)\phmr(t-x)\, , 
\label{r}\\
\phl(t,x) = \theta(x) \phpl (t+x) + \theta(-x)\phml(t+x)\, ,
\label{l}
\eea
where 
\be
\phpmr(\xi ) = \int_{0}^{+\infty} \frac{dk}{\pi \sqrt
{2k}}
\left[a^{\ast \pm}(k) \e^{ik\xi} +
a_\pm (k) \e^{-ik\xi}\right ] \,  , 
\label{pmr}
\ee
\be
\phpml (\xi) = \int_{0}^{+\infty} 
\frac{dk}{\pi \sqrt {2k}} 
\left[a^{\ast \pm}(-k) \e^{ik\xi} +
a_\pm (-k) \e^{-ik\xi}\right ] \,  . 
\label{pml}
\ee 
\vskip 1truecm 
\setlength{\unitlength}{1,25mm}
\begin{picture}(20,20)(-25,20) 
%x
\put(50,-3){\makebox(20,20)[t]{$x$}}
\put(28,20){\makebox(20,20)[t]{$t$}}
\put(25.2,0.7){\makebox(20,20)[t]{$\bullet$}}
\put(15.5,15){\makebox(20,20)[t]{$\phml$}}
\put(34.5,15){\makebox(20,20)[t]{$\phpr$}}
\put(34.5,-12){\makebox(20,20)[t]{$\phpl$}}
\put(16.5,-12){\makebox(20,20)[t]{$\phmr$}}
%lines
\qbezier(15,40)(35,20)(55,0)
\qbezier(55,40)(35,20)(15,0)
\thicklines
\qbezier(10,20)(30,20)(60,20)
\qbezier(35,0)(35,20)(35,40)
%direction
\put(61,20){\vector(1,0){0}}
\put(35,41){\vector(0,1){0}}
\end{picture}
\vskip 3 truecm
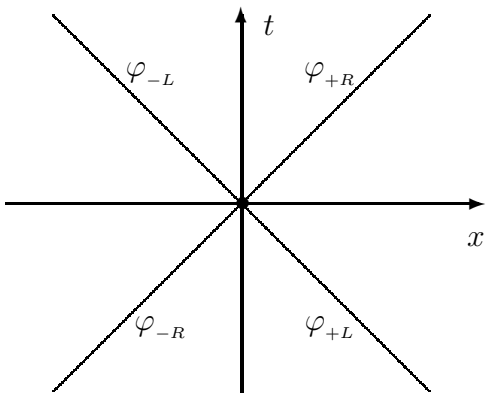
\centerline{Fig. 1. The localization of $\phpmr$ and $\phpml$ on the light cone.}
\bigskip 
\bigskip 

The four components $\phpmr$ and $\phpml$, whose localization 
is displayed on Fig. 1, couple each other through the defect at $x=0$. 
This characteristic feature of our system is captured by the 
correlation functions of $\phpmr$ and $\phpml$, 
we are going to derive now.  
Using (\ref{ccr3}) and the fact that $a_\pm(k)$ 
annihilate the Fock vacuum, 
one gets the following two-point functions 
\bea 
\langle \phpr (\xi_1) \phpr (\xi_2)\rangle = 
\langle \phmr (\xi_1) \phmr (\xi_2)\rangle = \nonumber \\
\langle \phpl (\xi_1) \phpl (\xi_2)\rangle = 
\langle \phml (\xi_1) \phml (\xi_2)\rangle = \nonumber \\ 
= \int_{0}^{+\infty} \frac{dk}{\pi }(k^{-1})_{\mu_0} 
\e^{-ik\xi_{12}} \, , \qquad \xi_{12} \equiv \xi_1-\xi_2\, ,  
\label{cf1}
\eea 
$(k^{-1})_{\mu_0}$ being the distribution \cite{Liguori:1997vd} 
\be 
(k^{-1})_{\mu_0} = \frac{d}{dk} \ln\frac{k}{\mu_0} \, .  
\label{distr1}
\ee
The derivative here is understood in the sense of distributions and 
$\mu_0$ is a free parameter with dimension of mass having a 
well-known (see e.g. \cite{Grignani:1988fx}) infrared origin. 
We observe for further use that the identity 
\be 
k\, (k^{-1})_{\mu_0} = 1 
\label{distr2}
\ee 
holds on $\RR_+$ and that 
\be
\int_{0}^{+\infty} \frac{dk}{\pi }(k^{-1})_{\mu_0} \e^{-ik\xi} 
= u(\mu \xi)\, , \qquad \mu \equiv \mu_0 \e^{\gamma_E} \, , 
\label{ft}
\ee 
where 
\be 
u(\xi)=-\frac{1}{\pi} \ln (|\xi|) -\frac{i}{2}\varepsilon (\xi) = 
-\frac{1}{\pi} \ln (i\xi + \epsilon )\, , \quad \epsilon > 0\, , 
\label{log}
\ee 
and $\gamma_E$ is the Euler constant. 
The correlators (\ref{cf1}) do not depend on the defect 
and coincide with the familiar defect-free ones. 
This conclusion obviously holds also for the commutators 
\be
[\phpr(\xi_1)\, ,\, \phpr(\xi_2)] = [\phmr(\xi_1)\, ,\, \phmr(\xi_2)] 
= -i\varepsilon (\xi_{12}) \, , 
\label{comm1}
\ee
\be
[\phpl(\xi_1)\, ,\, \phpl(\xi_2)] = [\phml(\xi_1)\, ,\, \phml(\xi_2)] \\
= -i\varepsilon (\xi_{12}) \, , 
\label{comm2}
\ee 
which follow directly from eqs. (\ref{cf1},\ref{ft},\ref{log}). 

The defect shows up in the mixed correlation functions in the following way. 
The transmission relates the plus and minus components with the 
same chirality:  
\be 
\langle \phpr (\xi_1) \phmr (\xi_2)\rangle = 
\int_{0}^{+\infty} \frac{dk}{\pi }(k^{-1})_{\mu_0} 
\e^{-ik\xi_{12}} T_+^-(k) \, ,
\label{cf2}
\ee 
\be 
\langle \phmr (\xi_1) \phpr (\xi_2)\rangle = 
\int_{0}^{+\infty} \frac{dk}{\pi }(k^{-1})_{\mu_0} 
\e^{-ik\xi_{12}} T_-^+(k) \, ,
\label{cf3}
\ee 
\be 
\langle \phpl (\xi_1) \phml (\xi_2)\rangle = 
\int_{0}^{+\infty} \frac{dk}{\pi }(k^{-1})_{\mu_0} 
\e^{-ik\xi_{12}} T_+^-(-k) \, ,
\label{cf4}
\ee 
\be 
\langle \phml (\xi_1) \phpl (\xi_2)\rangle = 
\int_{0}^{+\infty} \frac{dk}{\pi }(k^{-1})_{\mu_0} 
\e^{-ik\xi_{12}} T_-^+(-k) \, .
\label{cf5}
\ee 
The reflection instead relates different chiralities on the same half-line 
according to 
\be 
\langle \phpr (\xi_1) \phpl (\xi_2)\rangle = 
\int_{0}^{+\infty} \frac{dk}{\pi }(k^{-1})_{\mu_0} 
\e^{-ik\xi_{12}} R_+^+(k) \, ,
\label{cf6}
\ee 
\be 
\langle \phmr (\xi_1) \phml (\xi_2)\rangle = 
\int_{0}^{+\infty} \frac{dk}{\pi }(k^{-1})_{\mu_0} 
\e^{-ik\xi_{12}} R_-^-(k) \, ,
\label{cf7}
\ee 
\be 
\langle \phpl (\xi_1) \phpr (\xi_2)\rangle = 
\int_{0}^{+\infty} \frac{dk}{\pi }(k^{-1})_{\mu_0} 
\e^{-ik\xi_{12}} R_+^+(-k) \, ,
\label{cf8}
\ee 
\be 
\langle \phml (\xi_1) \phmr (\xi_2)\rangle = 
\int_{0}^{+\infty} \frac{dk}{\pi }(k^{-1})_{\mu_0} 
\e^{-ik\xi_{12}} R_-^-(-k) \, .
\label{cf9}
\ee 
{}Finally, 
\bea 
\langle \phpr (\xi_1) \phml (\xi_2)\rangle &=& 
\langle \phml (\xi_1) \phpr (\xi_2)\rangle = \nonumber \\
\langle \phmr (\xi_1) \phpl (\xi_2)\rangle &=&
\langle \phpl (\xi_1) \phmr (\xi_2)\rangle = 0 \, .   
\label{cf10}
\eea 
The plus-minus and left-right mixing is captured by (\ref{cf2}-\ref{cf5}) and 
(\ref{cf6}-\ref{cf9}) respectively and is a direct 
consequence of the impurity. In a simpler form this phenomenon appears 
also in the case of boundary conformal field theory \cite{Cardy:1984bb}. 

The integral representations (\ref{cf2}-\ref{cf9}) determine well-defined 
distributions allowing to analyze the locality properties of $\{\ph,\, \phd\}$. 
The explicit form of the commutators at generic points $t_1, x_1$ and 
$t_2, x_2$  is quite involved. Fortunately however it drastically simplifies 
at space-like separated points. In fact, in the domain $t_{12}^2-x_{12}^2 <0$ 
one finds 
\be
[\ph (t_1,x_1)\,  ,\, \ph (t_2,x_2)]= [\phd (t_1,x_1)\,  ,\, \phd (t_2,x_2)]=0 \, , 
\label{loc1} 
\ee
\be 
[\ph (t_1,x_1)\,  ,\, \phd (t_2,x_2)]= 
\frac{i}{2}[\varepsilon(x_{12})+\varepsilon(\widetilde {x}_{12})] 
\theta(x_1x_2) \, , 
\label{loc2} 
\ee 
where $\widetilde{x}_{12} \equiv x_1+x_2$. Therefore, like in the case without 
defects, $\ph$ and $\phd$ are {\it local} fields, but {\it not relatively local}. 
As recognized already in the early sixties \cite{bos}, this feature is the corner stone of 
bosonization. We will make essential use of it in the next section, establishing 
the statistics of the vertex operators and constructing, in particular, fermions from 
bosons. 

The symmetry properties of our system are strongly influenced by the 
impurity, which breaks down the 1+1 dimensional Poincar\'e group except 
of the invariance under time translations. Therefore the energy is conserved 
in accordance with the fact that our defects do not dissipate. 

Concerning the internal symmetries, we introduce the charges 
\be 
Q_{\epsilon Z} \equiv 
\int_{-\infty}^{+\infty}d\xi\, \prt_\xi \, \varphi_{\epsilon Z}(\xi) \, , 
\qquad  \epsilon = \pm\, ,\quad Z=R,\, L\, , 
\label{charges}
\ee 
which are the building blocks for constructing the Klein factors \cite{Goddard:1986bp} 
used in the vertex construction of affine Kac-Moody algebras.  
By definition $Q_{\epsilon Z}$ depend only on the asymptotic 
behavior of $\varphi_{\epsilon Z}(\xi)$ at $\xi=\pm \infty$. 
Using the correlation functions (\ref{cf1},\ref{cf2}-\ref{cf9}) one finds 
\be
\left [Q_{\epsilon_1 R} \, , \varphi_{\epsilon_2 R}(\xi) \right ] = 
-\left [Q_{\epsilon_1 R} \, , \varphi_{\epsilon_2 L}(\xi) \right ] 
= -i\delta_{\epsilon_1\epsilon_2} \, , 
\label{K1}
\ee
\be
\left [Q_{\epsilon_1 L} \, , \varphi_{\epsilon_2 R}(\xi) \right ] = 
-\left [Q_{\epsilon_1 L} \, , \varphi_{\epsilon_2 L}(\xi) \right ] 
= -i\delta_{\epsilon_1\epsilon_2} \, .   
\label{K2}
\ee

Summarizing, the defect divides each left and right branches 
$C_{{}_R}$ and $C_{{}_L}$ of the light cone $C=C_{{}_R}\cup C_{{}_L}$ 
in two components $C_{{}_{\pm R}}$ and $C_{{}_{\pm L}}$, where the chiral fields 
$\phpmr$ and $\phpml$ are localized. These fields are not independent: 
the reflection and transmission coefficients define a specific interaction 
between the left-right and plus-minus  components respectively. In the next subsection we
describe two  concrete sets of parameters $\{a,\, b,\, c,\, d\}$, which nicely illustrate 
both the above general structure and the characteristic features of 
bosonization with defects.

\subsection{Examples:} 

\begin{itemize}
\item{} {\bf quasi-conformal defects;}
\end{itemize}

We start by considering the one-parameter family of defects 
\begin{equation} 
\{a=1/\lambda,\, 0,\, 0,\, d=\lambda \not= 0\}\, . 
\label{qc0}
\end{equation} 
Since the dimensional parameters $b$ and $c$ are set to $0$, we call them  
quasi-conformal defects. These defects coincide with the 
permeable conformal walls introduced in 
\cite{Bachas:2001vj, DeWolfe:2001pq}. From (\ref{rcoef},\ref{tcoef})  one gets 
\be 
R_+^+(k) = -R_-^-(k) = r(\lambda)\equiv \frac{1-\lambda^2}{1+\lambda^2}\, , \qquad 
T_+^-(k) = T_-^+(k) = 1- r(\lambda) \, . 
\label{qc1}
\ee
Accordingly, one has in addition to (\ref{cf1}) the following 
non-trivial correlation functions 
\bea 
\langle \phpr (\xi_1) \phmr (\xi_2)\rangle &=& 
\langle \phmr (\xi_1) \phpr (\xi_2)\rangle = 
\nonumber \\
\langle \phpl (\xi_1) \phml (\xi_2)\rangle &=& 
\langle \phml (\xi_1) \phpl (\xi_2)\rangle = 
[1- r(\lambda)]\, u(\mu \xi_{12})\, , 
\label{qc2}
\eea 
\bea 
\langle \phpr (\xi_1) \phpl (\xi_2)\rangle &=& 
-\langle \phmr (\xi_1) \phml (\xi_2)\rangle = 
\nonumber \\
\langle \phpl (\xi_1) \phpr (\xi_2)\rangle &=& 
-\langle \phml (\xi_1) \phmr (\xi_2)\rangle = 
r(\lambda)\, u(\mu \xi_{12})\, , 
\label{qc3}
\eea 
which vanish in the conformal case. All correlators 
(\ref{cf1},\ref{ft},\ref{log},\ref{qc2},\ref{qc3})
of the quasi-conformal defect are expressed 
in terms of the logarithm $u(\mu\xi)$ and the parameter $\lambda$. 
In addition to the universal (defect independent) commutators 
(\ref{comm1},\ref{comm2}) one has: 
\be
[\phpr(\xi_1)\, ,\, \phmr(\xi_2)] = [\phml(\xi_1)\, ,\, \phpl(\xi_2)] 
= -i[1- r(\lambda)]\, \varepsilon (\xi_{12})\, , 
\label{qc4} 
\ee
\be
[\phpr(\xi_1)\, ,\, \phpl(\xi_2)] = -[\phml(\xi_1)\, ,\, \phmr(\xi_2)] \\
= -i\, r(\lambda)\, \varepsilon (\xi_{12})\, . \quad \; 
\label{qc5}
\ee 
\bigskip 

\begin{itemize} 
\item{} {\bf $\delta$-defects;} 
\end{itemize}

As a second example we consider the impurities defined by 
\be 
\{a=d=1,\, b=0,\, c=2\eta >0\}\, . 
\label{delta0} 
\ee
One usually refers to this one-parameter family as $\delta$-defects, 
because they can be implemented by coupling 
$\ph$ to the external potential $U(x) = 2\eta \delta(x)$. 
The reflection and  transmission
coefficients take the form 
\bea
R_+^+(k) = \frac{-i\eta}{k + i\eta} \, , \qquad 
R_-^-(k) = \frac{i\eta}{k - i\eta} \, ,
\label{rdelta} \\
T_+^-(k) = \frac{k}{k + i\eta}\, , \qquad 
T_-^+(k) = \frac{k}{k - i\eta}\, . 
\label{tdelta}
\eea 
It is worth mentioning that 
\be 
\lim_{k\to \infty} R_+^+(k) = \lim_{k\to \infty} R_-^-(k) = 0 \, , 
\label{deltalim}
\ee
which is actually an exclusive feature of the $\delta$-defects. In fact, 
it follows from (\ref{rcoef},\ref{tcoef}) that (\ref{rdelta},\ref{tdelta}) 
define the most general defects satisfying (\ref{deltalim}), which implies in turn  
that the correlation functions (\ref{cf6}-\ref{cf9}) are regular for
$\xi_1=\xi_2$. Indeed,  inserting (\ref{rdelta},\ref{tdelta}) in
(\ref{cf2}-\ref{cf9}) and performing the  integration over $k$ 
one gets the following explicit
two-point functions   
\be 
\langle \phpr (\xi_1) \phmr (\xi_2)\rangle = 
\langle \phml (\xi_1) \phpl (\xi_2)\rangle = v_-(\eta \xi_{12})\, ,
\label{deltacf1}
\ee 
\be 
\langle \phmr (\xi_1) \phpr (\xi_2)\rangle = 
\langle \phpl (\xi_1) \phml (\xi_2)\rangle = v_+(-\eta\xi_{12})\, ,
\label{deltacf2}
\ee 
\be 
\langle \phpr (\xi_1) \phpl (\xi_2)\rangle = 
\langle \phml (\xi_1) \phmr (\xi_2)\rangle = 
v_-(\eta\xi_{12}) - u(\mu\xi_{12})\, , 
\label{deltacf3}
\ee 
\be 
\langle \phmr (\xi_1) \phml (\xi_2)\rangle = 
\langle \phpl (\xi_1) \phpr (\xi_2)\rangle = 
v_+(-\eta\xi_{12}) - u(\mu\xi_{12})\, , 
\label{deltacf4}
\ee 
where $u$ is defined by (\ref{log}) and 
\be 
v_\pm(\xi) \equiv -\frac{1}{\pi} \e^{-\xi}
\, \Ei (\xi \pm i\epsilon) \, , 
\qquad \epsilon >0\, , 
\label{ei1}
\ee
$\Ei$ being the exponential-integral function. Recalling the expansion 
\be 
\Ei (\xi \pm i\epsilon) = \gamma_E + \ln (\xi \pm i\epsilon) + 
\sum_{n=1}^\infty \frac{\xi^n}{n\cdot n!}\, , 
\label{ei2}
\ee 
we see that $u(\xi)$ and $v_\pm(\mp \xi)$ have the same logarithmic 
singularity in $\xi=0$, confirming that the correlators 
(\ref{deltacf3},\ref{deltacf4}) are not singular at $\xi_1=\xi_2$. 

{}From (\ref{deltacf1}-\ref{deltacf4}) one gets the commutators: 
\be
[\phpr(\xi_1)\, ,\, \phmr(\xi_2)] = [\phml(\xi_1)\, ,\, \phpl(\xi_2)] 
= -2i\theta (\xi_{12}) \e^{-\eta\xi_{12}}\, , \qquad \qquad  
\label{comm3}
\ee
\be  
[\phpr(\xi_1)\, ,\, \phpl(\xi_2)] = [\phml(\xi_1)\, ,\, \phmr(\xi_2)] \\
= -2i\theta (\xi_{12}) \e^{-\eta\xi_{12}}+ i\varepsilon (\xi_{12})\, .
\label{comm4}
\ee 

We stress that in deriving (\ref{deltacf1}-\ref{deltacf4}) we essentially used that 
$\eta >0$. The correlators (\ref{deltacf1}-\ref{deltacf4}) 
are singular in the limit $\eta \to 0$, which forbids to recover from 
them the free case $\eta=0$. Such type of discontinuity 
appears \cite{Liguori:1997vd} also on the half-line between the 
scalar field quantized with Robin and Neumann boundary conditions.

\bigskip 
\sect{Vertex operators in presence of defects} 

We have enough background at this point for constructing vertex operators. For 
any couple $\zeta \equiv (\alpha ,\beta )\in \RR^2$ we introduce the field 
\be
V(t,x;\zeta ) = :\exp [i\sqrt \pi (\alpha \ph + \beta \phd )]: (t,x) \, , 
\label{v1} 
\ee
where the normal ordering $\, : \quad  :\, $ is taken with respect to 
the creation and annihilation operators $\{a^{\ast \pm}(k),\,  a_\pm (k)\}$. 
Like in the case without defect, the operators (\ref{v1}) generate an  
algebra $\cal V$. The exchange properties of the vertex operators $V(t,x;\zeta )$ 
determine their statistics. A standard calculation shows that 
\be 
V(t_1,x_1;\zeta_1 ) V(t_2,x_2;\zeta_2 ) = 
\E (t_{12},x_1,x_2;\zeta_1, \zeta_2 )\, 
V(t_2,x_2;\zeta_2 ) V(t_1,x_1;\zeta_1 ) 
\, , 
\label{exch}
\ee
the exchange factor $\E$ being  
\be 
\E(t_{12}, x_1, x_2;\zeta_1, \zeta_2 ) = 
\e^{-\pi \left [\alpha_1\ph (t_1,x_1) + \beta_1\phd (t_1,x_1)\, ,\, 
\alpha_2\ph (t_2,x_2) + \beta_2\phd (t_2,x_2)\right ]}\, .  
\label{exchf1}
\ee 
The statistics of $V(t,x;\zeta )$ is determined by the 
value of (\ref{exchf1}) at space-like distances $t_{12}^2-x_{12}^2 <0$. 
By means of (\ref{loc1},\ref{loc2}) one finds in this domain 
\be 
\E(t_{12}, x_1, x_2;\zeta_1, \zeta_2 ) = 
\e^{\frac{i\pi}{2}[(\alpha_1\beta_2+\alpha_2\beta_1)\varepsilon(x_{12}) 
+(\alpha_1\beta_2-\alpha_2\beta_1)\varepsilon({\widetilde x}_{12})] 
\theta (x_1 x_2)}\, .   
\label{exchf2}
\ee
Setting $\zeta_1=\zeta_2\equiv \zeta$ in (\ref{exchf2}) one 
obtains 
\be 
\E(t_{12}, x_1, x_2;\zeta, \zeta ) = 
\e^{i\pi \alpha \beta \varepsilon(x_{12}) \theta (x_1 x_2)}\, ,    
\label{exchf3}
\ee
which governs the statistics of $V(t,x;\zeta )$. 

It follows from 
(\ref{exchf3}) that the exchange properties of the vertex operators 
depend not only on the parameters $(\alpha,\, \beta)$, but also on the 
position. This is a new phenomenon in the context of bosonization, which has 
its origin in the breakdown of translation invariance by the impurity. 
The $\theta$-factor in the exponent of 
(\ref{exchf3}) implies that two vertex operators localized on 
the opposite sides of the impurity are exchanged as bosons. However, 
when the vertex operators are localized on the same half-line, 
they behave as anyons with statistics parameter 
\be 
\vartheta \equiv \alpha \beta \, . 
\label{statparam}
\ee 
For $\vartheta = 2k$ and $\vartheta = 2k+1$ with $k\in \ZZ$ one recovers 
Bose and Fermi statistics respectively. The remaining values of $\vartheta$ 
lead to abelian braid (anyon) statistics. 

It is instructive to consider the vertex algebras ${\cal V}_{\pm R}$ and 
${\cal V}_{\pm L}$ generated by 
\be 
V(t,x \gtrless 0;(\alpha,\alpha)) = :\exp [i\sqrt {\pi}\alpha \phpmr ]: (t-x)  
\equiv V_{{}_{\pm R}}(t-x;\alpha) \, , 
\label{vpmr}
\ee
\be 
V(t,x \gtrless 0;(\alpha,-\alpha)) = :\exp [i\sqrt {\pi}\alpha \phpml ]: (t+x)  
\equiv V_{{}_{\pm L}}(t+x;\alpha) \, , 
\label{vpml}
\ee
respectively. These vertex operators are localized on the branches 
$C_{{}_{\pm R}}$ and $C_{{}_{\pm L}}$ of the light cone. Up to 
unessential $\mu$-dependent multiplicative factor the 
universal (defect independent) vertex correlators are 
\bea 
\langle V_{{}_{+R}}(\xi_1;\alpha) V^*_{{}_{+R}}(\xi_2;\alpha) \rangle 
&=& \langle V_{{}_{+R}}(\xi_1;\alpha) V^*_{{}_{+R}}(\xi_2;\alpha) \rangle 
\nonumber \\ 
\langle V_{{}_{+L}}(\xi_1;\alpha) V^*_{{}_{+L}}(\xi_2;\alpha) \rangle 
&=& \langle V_{{}_{+L}}(\xi_1;\alpha) V^*_{{}_{+L}}(\xi_2;\alpha) \rangle 
\sim (\xi_{12}-i\epsilon)^{-\alpha^2} \, . 
\label{vc1}
\eea 
As expected, (\ref{vc1}) is a homogeneous function of degree $-\alpha^2$. 

Due to the interaction with the defect, there exist also non-trivial mixed 
vertex correlation functions, depending on the parameters $\{a,\, b,\, c,\, d\}$. 
The quasi-conformal defects (\ref{qc0}) lead for instance to 
\bea 
\langle V_{{}_{+R}}(\xi_1;\alpha) V^*_{{}_{-R}}(\xi_2;\alpha) \rangle 
&=& \langle V_{{}_{-R}}(\xi_1;\alpha) V^*_{{}_{+R}}(\xi_2;\alpha) \rangle =
\nonumber \\ 
\langle V_{{}_{+L}}(\xi_1;\alpha) V^*_{{}_{-L}}(\xi_2;\alpha) \rangle 
&=& \langle V_{{}_{-L}}(\xi_1;\alpha) V^*_{{}_{+L}}(\xi_2;\alpha) \rangle 
\sim (\xi_{12}-i\epsilon)^{-\frac{2\lambda^2}{1+\lambda^2}\alpha^2} \, ,  
\label{vc2}
\eea 
\bea 
\langle V_{{}_{+R}}(\xi_1;\alpha) V^*_{{}_{+L}}(\xi_2;\alpha) \rangle 
&=& \langle V_{{}_{-R}}(\xi_1;\alpha) V^*_{{}_{-L}}(\xi_2;\alpha) \rangle = 
\nonumber \\ 
\langle V_{{}_{+L}}(\xi_1;\alpha) V^*_{{}_{+R}}(\xi_2;\alpha) \rangle 
&=& \langle V_{{}_{-L}}(\xi_1;\alpha) V^*_{{}_{-R}}(\xi_2;\alpha) \rangle 
\sim (\xi_{12}-i\epsilon)^{-\frac{1-\lambda^2}{1+\lambda^2}\alpha^2} \, . 
\label{vc3}
\eea 
One has still homogeneous functions, whose degree is however  
$\lambda$-dependent. For impurities involving non-vanishing 
dimensional parameters $b$ and/or $c$ the correlation 
functions (\ref{vc2},\ref{vc3}) are no longer homogeneous functions of 
$\xi_{12}$. For the $\delta$-defect (\ref{delta0}) one gets for instance 
\bea 
\langle V_{{}_{+R}}(\xi_1;\alpha) V^*_{{}_{-R}}(\xi_2;\alpha) \rangle 
&=& \langle V_{{}_{-R}}(\xi_1;\alpha) V^*_{{}_{+R}}(\xi_2;\alpha) \rangle 
= \e^{\pi \alpha^2 v_-(\eta\xi_{12})} \, , 
\label{vc4}\\ 
\langle V_{{}_{+L}}(\xi_1;\alpha) V^*_{{}_{-L}}(\xi_2;\alpha) \rangle 
&=& \langle V_{{}_{-L}}(\xi_1;\alpha) V^*_{{}_{+L}}(\xi_2;\alpha) \rangle 
=\e^{\pi \alpha^2 v_+(-\eta\xi_{12})} \, ,  
\label{vc5}
\eea 
\bea 
\langle V_{{}_{+R}}(\xi_1;\alpha) V^*_{{}_{+L}}(\xi_2;\alpha) \rangle 
&=& \langle V_{{}_{-R}}(\xi_1;\alpha) V^*_{{}_{-L}}(\xi_2;\alpha) \rangle 
= \e^{\pi \alpha^2 [v_-(\eta\xi_{12})-u(\mu \xi_{12})]}\, ,
\label{cv6} \\ 
\langle V_{{}_{+L}}(\xi_1;\alpha) V^*_{{}_{+R}}(\xi_2;\alpha) \rangle 
&=& \langle V_{{}_{-L}}(\xi_1;\alpha) V^*_{{}_{-R}}(\xi_2;\alpha) \rangle 
=\e^{\pi \alpha^2 [v_+(-\eta\xi_{12})-u(\mu \xi_{12})]},  
\label{vc7}
\eea 
which contain the exponential-integral function (\ref{ei2}) and are not homogeneous. 

As a first application of the vertex algebra $\cal V$ for generic defect 
$\{a,\, b,\, c,\, d\}$ we consider the 
bosonization of the free massless Dirac field with impurity. 
In order to fix the notation, we first recall the massless Dirac equation
on $\RR \setminus \{0\}$. One has 
\be 
(\gamma_t \prt_t - \gamma_x \prt_x)\psi (t,x) = 0 \, , 
\qquad x\in \RR \setminus \{0\}\, , 
\label{deq} 
\ee 
where
\be 
\psi (t,x)=\pmatrix{ \psi_1(t,x) \cr \psi_2(t,x) \cr}\, ,  
\quad \qquad 
\gamma_t=\pmatrix{ 0 & 1 \cr 1 & 0 \cr}\, , \quad \qquad
\gamma_x=\pmatrix{ 0 & 1 \cr -1 & 0 \cr} \, .
\label{gamma}
\ee 
The standard vector and axial currents are 
\be 
j_\nu (t,x) = \overline \psi (t,x) \gamma_\nu \psi (t,x) \, , \quad  
\qquad j_\nu^5 (t,x) = \overline \psi (t,x) \gamma_\nu \gamma^5 \psi (t,x) \,  , 
\quad \qquad \nu = t,x \,  , 
\label{currents} 
\ee
with $\psi \equiv \psi^\ast \gamma_t $ and 
$\gamma^5 \equiv -\gamma_t\gamma_x $. From eq.(\ref{deq}) it follows 
that both $j_\nu $ and $j_\nu^5$ are conserved.  
Moreover, the $\gamma^5$-identities $\gamma_t\gamma^5 = -
\gamma_x $ and 
$\gamma_x \gamma^5 = - \gamma_t $ imply the relations 
\be
j_t^5 = - j_x \, ,\quad  \qquad j_x^5 = -j_t \, . 
\label{currrel}
\ee

Our main goal now is to quantize (\ref{deq}) in terms of 
the fields $\ph $ and $\phd $, establishing the defect 
boundary conditions on $\psi $ encoded in (\ref{bc}). For this 
purpose we set 
\be
\psi_1 (t,x) = \mu^{\alpha^2/2}V(t,x;\zeta_1 = (\alpha ,\alpha )) = 
\mu^{\alpha^2/2}:\exp(i\sqrt \pi \alpha \phr ): (t,x)   
\, , 
\label{psione} 
\ee
\be 
\psi_2 (t,x) = \mu^{\alpha^2/2}V(t,x;\zeta_2 = (-\alpha ,\alpha )) = 
\mu^{\alpha^2/2}:\exp(i\sqrt \pi \alpha \phl ): (t,x) \, , 
\label{psitwo} 
\ee 
where $\mu$ is the infrared scale introduced in (\ref{ft}). 

One easily verifies 
that $\psi $, defined by eqs. (\ref{psione},\ref{psitwo}), 
satisfies the Dirac equation on $\RR \setminus \{0\}$. 
Moreover, one has the anticommutation relations 
\be 
\psi_i (t_1,x_1) \psi_j (t_2,x_2) = 
- \psi_j (t_2,x_2) \psi_i (t_1,x_1) \, , 
\label{anticomm}
\ee  
{}for $|t_{12}| < |x_1 - x_2|$ and $x_1 x_2 >0$, provided that 
\be
\alpha^2 = 2k+1 \, , \qquad k \in \NN \, , 
\label{cond}
\ee
which is imposed to the end of this section.  

The next step is to construct the quantum currents (\ref{currents}). We 
adopt the point-splitting procedure, considering the limits 
\be
j_\nu (t,x) = {1\over 2} \lim_{\sigma \to \pm 0} Z(\sigma) 
\left [\, \overline \psi (t,x)\gamma_\nu \psi (t, x+\sigma ) + 
\overline \psi (t, x+\sigma ) \gamma_\nu \psi (t,x)\, \right ]\, , \qquad x \gtrless 0\, , 
\label{pointsplit1}  
\ee
where $Z(\sigma)$ implements the renormalization. 
The basic general formula for 
evaluating (\ref{pointsplit1}), is obtained by normal ordering the product 
$V^\ast (t,x+\sigma;\zeta ) V(t,x;\zeta )$. One has 
\bea
V^\ast (t,x+\sigma;\zeta ) V(t,x;\zeta ) &=& 
\nonumber \\ 
:\exp \Bigl \{ i\sqrt \pi \bigl [ \alpha \ph (t,x) - \alpha \ph (t,x+\sigma ) 
&+& \beta \phd (t,x) - \beta \phd (t,x+\sigma ) \bigr ] \Bigr \}: 
\nonumber \\
\exp \Bigl \{ {\pi \over 4} \bigl [ (\alpha + \beta )^2 
\langle \phr(t,x+\sigma) \phr(t,x)\rangle 
&+& (\alpha - \beta )^2\langle \phl(t,x+\sigma) \phl(t,x)\rangle 
\nonumber \\
+ (\alpha^2 - \beta^2 )\bigl (\langle \phr(t,x+\sigma) \phl(t,x)\rangle
&+& \langle \phl(t,x+\sigma) \phr(t,x)\rangle\bigr ) \bigr ] \Bigr \}\, . 
\label{VV}
\eea 
{}For the special values $\zeta_1=(\alpha,\, \alpha)$ and 
$\zeta_2 =(-\alpha,\, \alpha)$ the mixed $R-L$ and $L-R$ correlation functions 
drop out. Setting    
\be 
Z(\sigma ) = 
{-\sigma^{\alpha^2 -1}\over 2 \alpha \sin \left ({\pi\over 2}\alpha^2\right ) }
\, , 
\label{rc1} 
\ee 
and performing the limit in (\ref{pointsplit1}), 
one finds the conserved current 
\be 
j_\nu (t,x) = {\sqrt \pi}\, \prt_\nu \ph (t,x) \, .  
\label{jph} 
\ee 
Thus one recovers the same type of relation as in conventional 
bosonization \cite{bos} without impurities. 

In analogy with (\ref{pointsplit1}) we introduce the axial current by  
\be 
j_\nu^5 (t,x) = 
{1\over 2} \lim_{\sigma \to \pm 0} Z(\sigma) 
\left [\, \overline \psi (t,x)\gamma_\nu \gamma^5 \psi (t, x+\sigma ) + 
\overline \psi (t, x+\sigma ) \gamma_\nu \gamma^5 \psi (t,x)\, \right ]\, , 
\quad x \gtrless 0\, , 
\label{pointsplit2}  
\ee 
The vector current result and the $\gamma^5$-identities directly imply that 
the limit in the right hand side of (\ref{pointsplit2}) exists and 
\be 
j^5_\nu (t,x) = {\sqrt \pi}\, \prt_\nu \phd (t,x) \, . 
\label{jphd} 
\ee 
The classical relations (\ref{currrel}) are thus respected on quantum level 
as well. 

Eqs.(\ref{jph},\ref{jphd}) imply that the defect boundary conditions on 
$\psi $ at $x=0$ are most conveniently formulated in terms of the 
currents, which are the simplest observables of the fermion field. 
Combining  (\ref{bc}) with (\ref{jph},\ref{jphd}) one obtains 
\bea
\int_{+0}^{+\infty} dx\, j_x(t,x) = 
a \int_{-\infty}^{-0}dx\, j_x(t,x) + b\, j_x(t,-0) \, , 
\label{fbc1} \\ 
j_x(t,+0) = 
c \int_{-\infty}^{-0}dx\, j_x(t,x) + d\, j_x(t,-0) \, .  
\label{fbc2}
\eea 
The linear and local boundary conditions on $\ph$ are therefore translated in both 
non-linear and non-local conditions on $\psi$. 

Summarizing, we have shown that vertex operators 
in the presence of a point-like defect admit $x$-dependent 
anyon statistics.  Afterwards we established a bosonization procedure for 
the free massless Dirac field in $\RR \setminus {0} $. The relative 
vector and axial currents have been expressed in terms of 
$\ph$ and $\phd$ respectively. Taking as an example the massless 
Thirring model with defect, we will extend in the next section 
the bosonization procedure to the case of current-current interactions. 

\bigskip 
\sect{Thirring model with defect} 

We will first solve the massless Thirring model \cite{Thirr} with a $\delta$-defect 
(\ref{delta0}), generalizing afterwards the solution to a generic point-like 
defect $\{a,\, b,\, c,\, d\}$. 
The classical dynamics of the model is governed 
by the equation of motion 
\be 
i(\gamma_t \prt_t - \gamma_x \prt_x)\Psi (t,x) = 
g\left [\gamma_t J_t(t,x) - 
\gamma_x J_x (t,x) \right ] \Psi (t,x) 
\, , \quad x\not=0\, , 
\label{theqm}
\ee 
where $g\in \RR $ is the coupling constant and $J_\nu$ is 
the conserved current 
\be 
J_\nu (t,x) = \overline \Psi (t,x) \gamma_\nu \Psi (t,x) \, ,  
\label{thcurr}
\ee 
which, according to (\ref{fbc1},\ref{fbc2}), satisfies the defect boundary conditions 
\bea
\int_{+0}^{+\infty} dx\, J_x(t,x) &=& 
\int_{-\infty}^{-0}dx\, J_x(t,x)  \, , 
\label{thfbc1} \\ 
J_x(t,+0) - J_x(t,-0)&=& 
2\eta \int_{-\infty}^{-0}dx\, J_x(t,x)  \, .  
\label{thfbc2}
\eea 
{}For quantizing the system (\ref{theqm}-\ref{thfbc2}), we introduce the fields 
\be
\Psi_1 (t,x) = \mu^\gamma \, V(t,x;\zeta_1 = (\alpha,\beta )) = 
\mu^\gamma :\exp[i\sqrt \pi (\alpha \ph + \beta \phd )]: (t,x)  \, , 
\label{psi1} 
\ee
\be
\Psi_2 (t,x) = \mu^\gamma \, V(t,x;\zeta_2 = (\alpha ,-\beta )) = 
\mu^\gamma :\exp[i\sqrt \pi (\alpha \ph - \beta \phd )]: (t,x) \, ,  
\label{psi2} 
\ee 
where 
\be 
\gamma \equiv \frac{\alpha^2+\beta^2}{2} >0\, , 
\label{gamma}
\ee
and the correlation functions of $\{\ph,\, \phd\}$ 
are fixed by (\ref{deltacf1}-\ref{deltacf4}). Moreover, we require 
\be 
\alpha \beta = 2k+1\, , \qquad k\in \ZZ\, , 
\label{fermistat}
\ee
which according to the results of the previous section ensures Fermi 
statistics when $\Psi_{1,2}$ are localized at the same side 
of the defect. 

The quantum current $J_\nu $ is constructed 
in analogy with (\ref{pointsplit1}), setting  
\be
J_\nu (t,x) = {1\over 2} \lim_{\sigma \to \pm 0} Z_\nu (x;\sigma) 
\left [\, \overline \Psi (t,x)\gamma_\nu \Psi (t, x+\sigma ) + 
\overline \Psi (t, x+\sigma ) \gamma_\nu \Psi (t,x)\, \right ]\, , 
\quad x \gtrless 0\, ,  
\label{pointsplit3}  
\ee
without summation over $\nu $. The presence of two $x$-dependent 
renormalization constants $Z_t(x;\sigma)$ and $Z_x(x;\sigma)$ 
in (\ref{pointsplit3}) is a consequence of the fact that both 
translation and Lorentz invariance are broken by the defect. 
In order to satisfy the defect boundary conditions 
(\ref{thfbc1},\ref{thfbc2}), we shall determine 
$Z_t(x;\sigma)$ and $Z_x(x;\sigma)$ in such a way that 
\be 
J_\nu (t,x) = {\sqrt \pi }\, \prt_\nu \ph (t,x) \, .  
\label{thvcurrent}
\ee 
{}For this purpose we first observe that the operator products 
under the limit (\ref{pointsplit3}) have the following expansions 
for $\sigma \to 0$: 
\bea
{1\over 2}\left [\Psi_1^*(t,x+\sigma ) \Psi_1 (t,x) + 
\Psi_1^*(t,x) \Psi_1 (t,x+\sigma ) \right ] = 
\nonumber \\ 
\sigma^{1-\gamma} 
\left [\alpha \prt_x \ph (t,x) + \beta \prt_x \phd (t,x) 
+ O(\sigma^2 ) \right ]\chi (x;\alpha ,\beta ) \, , 
\label{psi1psi1} 
\eea 
\bea 
{1\over 2}\left [\Psi_2^*(t,x+\sigma) \Psi_2 (t,x) + 
\Psi_2^*(t,x) \Psi_2 (t,x+\sigma ) \right ] = 
\nonumber \\ 
\sigma^{1-\gamma}  
\left [\alpha \prt_x \ph (t,x) - \beta \prt_x \phd (t,x) 
+ O(\sigma^2 ) \right ]\, \chi (x;\alpha ,\beta ) \, , 
\label{psi2psi2}
\eea 
where $\chi(x;\alpha ,\beta )$ is a function given by 
\be 
\chi(x;\alpha ,\beta ) = 
\sqrt{\pi} \sin\left(\frac{\pi}{2}\alpha \beta\right) 
\e^{\left\{\frac{\pi}{4}(\alpha^2-\beta^2) \left [v_+(-2\eta x) - u(2\mu x) + 
v_-(-2\eta x)-u(-2\mu x)\right ]\right\}}\, .  
\label{chi1}
\ee
We conclude therefore that defining 
\be 
Z_t(x;\sigma) = \frac{-\sigma^{\gamma -1}\sqrt \pi}
{2 \beta \chi (x;\alpha ,\beta )}\, , \qquad  
Z_x(x;\sigma) = \frac{-\sigma^{\gamma -1}\sqrt \pi}
{2 \alpha \chi (x;\alpha ,\beta )}\, ,
\label{ZZ}
\ee
the limit (\ref{pointsplit3}) precisely reproduces (\ref{thvcurrent}). 

Let us turn now to the quantum version of the equation of motion. 
In view of (\ref{thvcurrent}), one gets from (\ref{theqm})
\be
i(\gamma_t \prt_t - \gamma_x \prt_x)\Psi (t,x) = 
g {\sqrt \pi } : \left (\gamma_t \prt_t \ph - 
\gamma_x \prt_x \ph \right ) \Psi : (t,x)  \, .  
\label{qeqm}
\ee
Now, using the explicit form (\ref{psi1},\ref{psi2}) of $\Psi$, one easily 
verifies that (\ref{qeqm}) is satisfied provided that 
\be 
\alpha - \beta = -g \, .
\label{condeqm}
\ee 
Combining eq. (\ref{fermistat}) and eq. (\ref{condeqm}), 
we obtain two families of solutions 
\be 
\alpha_{1,2} = -{g\over 2} \pm \sqrt {{g^2\over 4} + (2k+1) } 
\, , \qquad \beta_{1,2} = \alpha_{1,2} + g  \, ,
\label{sol} 
\ee
parameterized by $k \in \ZZ$ with the constraint 
\be 
2k+1 \geq -{g^2\over 4} \,  , 
\label{condsol} 
\ee
ensuring that $\alpha , \beta \in \RR$. The freedom associated with 
$k\in \ZZ$ is present also in the Thirring model without defect. 
Since Lorentz invariance is preserved there, it is natural to 
require in addition that the Lorentz spin of $\Psi $ takes 
the canonical value ${1\over 2}$, which fixes $k=0$. 

The above solution of the Thirring model has a straightforward generalization 
to a generic defect $\{a,\, b,\, c,\, d\}$. In that case the function 
$\chi(x;\alpha ,\beta )$ takes the form 
\be 
\chi(x;\alpha ,\beta ) = 
\sqrt{\pi} \sin\left(\frac{\pi}{2}\alpha \beta\right) 
\e^{\left\{\frac{\pi}{4}(\alpha^2-\beta^2) 
\left [\langle \phpl (x) \phpr (-x)\rangle  + 
\langle \phpr (-x) \phpl (x)\rangle \right ]\right\}}\, .  
\label{chi2}
\ee 
It is easily seen that the corresponding change in 
the renormalization constants $Z_\nu$ 
does not affect the values (\ref{sol}) of the parameters 
$\alpha $ and $\beta$ as functions of the coupling constant 
$g$. 

\bigskip 
\sect{Non-abelian bosonization} 

{}Following the Frenkel-Kac construction \cite{Goddard:1986bp} 
of the vertex representation of 
the affine Kac-Moody algebra $\widehat {sl}(2)$, we introduce the operators 
\be 
H_{\epsilon Z}(\xi) = {\sqrt \pi} \prt_\xi \varphi_{\epsilon Z}(\xi) \, , 
\qquad 
E^\pm_{\epsilon Z}(\xi) = 
\mu :\e^{\pm i \sqrt {2\pi}  \varphi_{\epsilon Z}(\xi)}: \, , 
\label{km1} 
\ee 
where $\epsilon = \pm,\,  Z=L,R$. Using eqs. (\ref{comm1},\ref{comm2}), 
{\it for fixed} $\{\epsilon,\, Z\}$ one gets the well-known $\widehat {sl}(2)$ 
commutation relations: 
\be
[H_{\epsilon Z}(\xi_1)\, ,\, H_{\epsilon Z}(\xi_2)] = 2\pi i\, \delta^\prime (\xi_{12})\II\, , 
\qquad \qquad \qquad \qquad  \; \; 
\label{hh} 
\ee 
\be
[H_{\epsilon Z}(\xi_1)\, ,\, E^\pm_{\epsilon Z}(\xi_2)] = \pm 2\pi \, \delta (\xi_{12}) 
\sqrt {2}\, E^\pm_{\epsilon Z}(\xi_2)\, , 
\qquad \quad \; \; \;  
\label{he}
\ee
\be 
[E^+_{\epsilon Z}(\xi_1)\, ,\, E^-_{\epsilon Z}(\xi_2)] = 2\pi i\, \delta^\prime (\xi_{12})\II  
+ 2\pi \, \delta (\xi_{12})\, H_{\epsilon Z}(\xi_1)\, , 
\label{epm}
\ee
\be 
[E^+_{\epsilon Z}(\xi_1)\, ,\, E^+_{\epsilon Z}(\xi_2)] = 
[E^-_{\epsilon Z}(\xi_1)\, ,\, E^-_{\epsilon Z}(\xi_2)] = 0 \, .
\qquad \; \;  \; \;  
\label{eppmm}
\ee 
In this way one recovers four vertex representations 
$\{\varrho_{\epsilon Z}\, :\, \epsilon =\pm,\, Z=R,\, L\}$ of 
$\widehat {sl}(2)$. This is not surprising because, as explained in section 2, 
the defect remains hidden when the theory is restricted on any of the components 
$C_{\epsilon Z}$ of the light cone. Keeping in mind that all four representations 
act in the Hilbert space where the vertex algebra $\cal V$ is represented, 
one can study also the interplay between the generators of different 
$\varrho_{\epsilon Z}$. Let us observe first of all that  
$\varrho_{+R}$ and $\varrho_{-L}$ as well as $\varrho_{-R}$ and 
$\varrho_{+L}$ commute because of (\ref{cf10}). However, 
since $\{\varphi_{\epsilon Z}\}$ interact among themselves through the defect, 
there is a non-trivial interplay among the other four pairs 
$\{\varrho_{+ R},\, \varrho_{+L}\}$,  
$\{\varrho_{- R},\, \varrho_{- L}\}$, 
$\{\varrho_{+ R},\, \varrho_{-R}\}$ and 
$\{\varrho_{+L},\, \varrho_{- L}\}$  
of representations. For a generic 
defect $\{a,\, b,\, c,\, d\}$ the commutator of two generators belonging 
to $\varrho_{\epsilon_1 Z_1}$ and $\varrho_{\epsilon_2 Z_2}$
is in general a {\it bilocal} operator of the type 
\be 
B_{\epsilon_1 Z_1,\, \epsilon_2 Z_2}^{\; \; \pm \quad \; \pm}(\xi_1,\xi_2)  \equiv \, 
:\e^{\pm i \sqrt {2\pi}  \varphi_{\epsilon_1 Z_1}(\xi_1) \pm i \sqrt {2\pi} 
\varphi_{\epsilon_2 Z_2}(\xi_2)}: \, . 
\label{bilocal}
\ee 
It turns out that the mixed commutators within the pairs 
$\{\varrho_{+ R},\, \varrho_{+L}\}$,  
$\{\varrho_{- R},\, \varrho_{- L}\}$, 
$\{\varrho_{+ R},\, \varrho_{-R}\}$ and 
$\{\varrho_{+L},\, \varrho_{- L}\}$ have all the same structure. So, let us 
consider for illustration the commutators between 
$\varrho_{+R}$ and $\varrho_{+L}$. One finds   
\be
[H_{+ R}(\xi_1)\, ,\, H_{+ L}(\xi_2)] = i\prt_{\xi_1}f(\xi_{12})\II\, ,   
\qquad \qquad \qquad \qquad  
\label{hrhl} 
\ee 
\be
[H_{+ R}(\xi_1)\, ,\, E^\pm_{+ L}(\xi_2)] = \pm f(\xi_{12})  
\sqrt {2}\, E^\pm_{+ L}(\xi_2)\, , 
\qquad \quad \; \; \; \;  
\label{hrel}
\ee
\be
[H_{+ L}(\xi_1)\, ,\, E^\pm_{+ R}(\xi_2)] = \mp f(-\xi_{12})
\sqrt {2}\, E^\pm_{+ R}(\xi_2)\, , 
\qquad \quad \;  
\label{hler}
\ee
\be 
[E^+_{+ R}(\xi_1)\, ,\, E^+_{+ L}(\xi_2)] = g_+(\xi_{12}) 
B_{+R,\, +L}^{\; \, + \; \; \; +}(\xi_1,\xi_2) \, , \quad \qquad 
\label{eprepl}
\ee
\be
[E^-_{+ R}(\xi_1)\, ,\, E^-_{+ L}(\xi_2)] =  g_+(\xi_{12}) 
B_{+R,\, +L}^{\; \, - \; \; \; -}(\xi_1,\xi_2) \, , 
\quad \qquad \; 
\label{emreml}
\ee
\be 
[E^+_{+R}(\xi_1)\, ,\, E^-_{+L}(\xi_2)] = g_-(\xi_{12}) 
B_{+R,\, +L}^{\; \, + \; \; \; -}(\xi_1,\xi_2) \, , 
\quad \qquad 
\label{epreml}
\ee
\be
[E^-_{+R}(\xi_1)\, ,\, E^+_{+L}(\xi_2)] = g_-(\xi_{12}) 
B_{+R,\, +L}^{\; \, - \; \; \; +}(\xi_1,\xi_2) \, , 
\quad \qquad \;  
\label{emreml}
\ee 
where $f$ and $g_\pm$ are some functions depending on the 
defect and thus on the parameters $\{a,\, b,\, c,\, d\}$. For 
the quasi-conformal defects (\ref{qc0}) one has  
\be 
f(\xi ) = 2\pi r(\lambda) \delta (\xi )\, , \qquad 
r(\lambda) = \frac{1-\lambda^2}{1+\lambda^2} \, , 
\label{fqc}
\ee
\be 
g_\pm(\xi )= \pm 2i\, \mu^{2\pm r(\lambda)}\, 
\sin[\pi r(\lambda)]\, \varepsilon (\xi)\, |\xi|^{\pm 2r(\lambda)}  \, .  
\label{gcq}
\ee
The $\delta$-defects (\ref{delta0}) lead instead to 
\be 
f(\xi ) = -2\pi\, \eta\, \theta(\xi) \e^{-\eta \xi}\, , 
\label{fd}
\ee
\be 
g_\pm(\xi )= \pm 2i\, \mu^2\, 
\sin \left (2\pi \e^{-\eta \xi}\right )\, \theta (\xi)\, \e^{\pm \gamma (\xi;\eta,\mu)} \, ,  
\label{gd}
\ee 
with 
\be 
\gamma (\xi;\eta,\mu) = 2\left [\e^{-\eta \xi}\left ( \gamma_E + \ln (\eta |\xi|) + 
\sum_{n=1}^\infty \frac {(\eta\xi)^n}{n\cdot n!}\right ) - \ln (\mu |\xi|) \right ] . 
\label{gdg}
\ee

The commutators (\ref{hrhl}-\ref{emreml}) deserve some comments. 
In analogy with (\ref{hh}), the commutator of the left and right 
Cartan generators is proportional to the identity operator $\II$. A first novelty is 
the central extension multiplication factor $i\prt_{\xi_1}f(\xi_{12})$, which is different 
and depends on the defect. The commutation 
of Cartan generators with step operators reproduces the latter up to a factor 
which is the integral of the central extension in (\ref{hrhl}). Finally, the
commutation of step operators leads,  up to the structure functions $g_\pm$, 
to the bilocal operators (\ref{bilocal}). 

It is perhaps useful to recall that the representations $\{\varrho_{\epsilon Z}\}$ of 
$\widehat {sl}(2)$ have a direct physical application. 
They describe the symmetry content of the $SU(2)$-invariant massless 
Thirring model with a $\delta$-impurity. Without impurity the model 
has been solved long ago with bosonization by M. Halpern \cite{Halpern:1975nm}. 
In the presence of a $\delta$-defect the solution is a direct 
generalization of our results in the previous section. 

Let us consider now the energy-momentum tensor $\Theta$ of the quantum field 
$\varphi$ interacting with the defect \cite{Mintchev:2004jy}. The chiral components 
\be 
\Theta_Z (x,\xi) = \theta(-x) \Theta_{- Z} (\xi) + \theta(x) \Theta_{+ Z} (\xi) 
\label{sug1}
\ee 
of $\Theta$ can be expressed in terms of the generators $H_{\epsilon Z}$ 
by means of 
\be 
\Theta_{\epsilon Z} (\xi) = \frac{1}{2\pi} :H_{\epsilon Z}\, H_{\epsilon Z}:(\xi) \, ,  
\label{sug2}
\ee 
which is precisely the Sugawara representation \cite{Goddard:1986bp}. 
As expected, {\it for fixed} $\{\epsilon,\, Z\}$ one finds 
\be 
[\Theta_{\epsilon Z} (\xi_1)\, ,\, \Theta_{\epsilon Z} (\xi_2)] = 
2 i \delta^{\, '} (\xi_{12}) \Theta_{\epsilon Z} (\xi_1) - 
\frac{i}{6\pi } \delta^{\, '''} (\xi_{12})\, \II \, . 
\label{sug3}
\ee 
{}From the properties of $H_{\epsilon Z}$ one infers that 
$\Theta_{+R}$ commutes with $\Theta_{-L}$ as well as  
$\Theta_{-R}$ commutes with  $\Theta_{+L}$. The remaining commutators 
are however non-trivial. In the quasi-conformal case one finds for instance 
\be 
[\Theta_{+R} (\xi_1)\, ,\, \Theta_{+L} (\xi_2)] = 
i \delta^{\, '} (\xi_{12}) r(\lambda) 
\left [\Theta_{+R,\, +L} (\xi_1) + \Theta_{+L,\, +R} (\xi_1) \right ] 
- \frac{ir(\lambda)^2}{6\pi } \delta^{\, '''} (\xi_{12})\II \, ,  
\label{sug4}
\ee 
where 
\be 
\Theta_{\epsilon_1 Z_1,\, \epsilon_2 Z_2} (\xi) = 
\frac{1}{2\pi} :H_{\epsilon_1 Z_1}\, H_{\epsilon_2 Z_2}:(\xi) \, .
\label{sug5} 
\ee
The appearance of {\it mixed} Sugawara terms of the type (\ref{sug5}) is a new feature, 
which has once more its origin in the left-right and plus-minus mixing due to the defect. 
We observe also that the commutator (\ref{sug4}) has a central term, the central 
charge being renormalized by a factor of $r(\lambda)^2$ with respect to 
(\ref{sug3}). One might be tempted to change the normalization of $\Theta_{\epsilon Z}$ 
in order to eliminate all factors $r(\lambda)$ from the right hand side of (\ref{sug4}), 
but then the inverse of this factor will appear in (\ref{sug3}). 

It is worth mentioning that the operators (\ref{sug2},\ref{sug5}) close
actually an  algebra. A straightforward but long computation using the RT algebra 
relations (\ref{ccr1}-\ref{ccr3}) gives in fact  
\be 
[\Theta_{+Z} (\xi_1)\, ,\, \Theta_{+R,\, +L} (\xi_2)] = 
i \delta^{\, '} (\xi_{12})  
\left [r(\lambda) \Theta_{+Z} (\xi_1) + \Theta_{+R,\, +L} (\xi_1) \right ] 
- \frac{i r(\lambda) }{6\pi }\delta^{\, '''} (\xi_{12})\II   
\label{sug6}
\ee 
and 
\bea 
[\Theta_{+R,\, +L} (\xi_1)\, ,\, \Theta_{+R,\, +L} (\xi_2)] = 
\qquad \qquad \qquad \qquad \quad 
\nonumber \\
i \delta^{\, '} (\xi_{12}) 
\left [\Theta_{+R} (\xi_1)+r(\lambda) \Theta_{+R,\, +L} (\xi_1) + 
\Theta_{+L} (\xi_1) \right ]  - 
\frac{i[r(\lambda)^2 +1]}{12\pi } \delta^{\, '''} (\xi_{12})\II \, ,  
\label{sug7}
\eea 
which complete the picture in the quasi-conformal case. 
Like in the Kac-Moody algebra, for more general 
defects the commutators (\ref{sug4},\ref{sug6},\ref{sug7}) 
involve bilocal operators of the form  
\be 
\Theta_{\epsilon_1 Z_1,\, \epsilon_2 Z_2} (\xi_1,\xi_2) = 
\frac{1}{2\pi} :H_{\epsilon_1 Z_1}(\xi_1)\, H_{\epsilon_2 Z_2}(\xi_2): \, .
\label{sug8} 
\ee 

Summarizing, we have shown in this section how some familiar structures from 
conformal field theory are modified by the presence of a point-like impurity, 
which preserves unitarity and locality. Together with the left-right mixing, 
a relevant characteristic feature is the appearance of bilocal operators.  

We conclude by observing that the 
above construction of the $\widehat {sl}(2)$ Kac-Moody algebra and 
the Sugawara representation of the energy-momentum tensor can be 
extended to the case of $\widehat {sl}(n)$ with $n>2$. The Klein factors, 
ensuring the right statistics of the Kac-Moody generators \cite{Goddard:1986bp}, 
are constructed in terms of the charges defined by (\ref{charges}).

\bigskip 
\sect{Conclusions and perspectives} 

Chiral fields, vertex operators and conformal field theory on the plane played in the past
two decades a fundamental role in the development of both theoretical physics and
mathematics. Considering the simplest case of a massless scalar quantum field 
$\varphi$, we propose in the present paper a generalization which consists in a theory 
on the plane without a line, where dissipationless boundary 
conditions are imposed on $\varphi$. In physical terms the line represents the 
world-line of a point-like defect. The interaction of the 
defect with $\varphi$ breaks down conformal 
invariance, but the breaking is fully under control and the theory is unitary. 
Dimensional parameters appear in the model and the left and right chiral sectors 
couple trough the defect. Although deformed, most of the basic structures 
(left and right chiral fields, vertex and Kac-Moody algebras, 
energy-momentum tensor,...) keep a 
well-defined physical and mathematical meaning. The vertex operators 
still carry anyon statistics and some of their correlators 
have anomalous dimension. Moreover, it turns out that bosonization 
can be successfully developed
in this context.  As an application of this method we solved explicitly 
the massless Thirring model with a generic point-like defect. 
We also constructed in this framework the vertex representation 
of the $\widehat {sl}(2)$ Kac-Moody algebra, establishing the  
interplay between the left and right sectors mediated by the defect. 
Using the Sugawara representation, we derived the main properties 
of energy-momentum tensor as well. The general results of our investigation have 
been illustrated on the concrete examples of quasi-conformal 
and $\delta$-defects.  

The method of bosonization with a dissipationless point-like defect, developed 
in this paper, suggest some new topics for research. 
An interesting issue, deserving further investigation,  
is the algebra generated by the energy-momentum tensor 
and the vertex operators. Concerning the physical applications, 
one can use our results for solving models
with impurities in condensed  matter physics. The simplest one 
is the generalization of the Luttinger
model to the case with  impurities. It will be also interesting to extend 
our framework to integrable systems with non-trivial bulk scattering. 
In this case the two-body bulk scattering matrix shows up \cite{Mintchev:2003ue} 
as a non-trivial exchange factor in the reflection-transmission algebra. 
Such a generalization is expected to produce a sort of quantum deformation 
\cite{Ragoucy:2001zy,Li} of the vertex algebra with defect.

\bigskip 
\bigskip

\noindent{\bf Acknowledgments} 
\bigskip 

\noindent It is a great pleasure to thank E. Pilon, C. Roger  and especially R. Stora for useful
discussions.  M. M. would also like to thank LAPTH in Annecy for the kind hospitality.

\end{document}